\documentstyle[preprint,floats,pra,aps,epsfig]{revtex} 
\tightenlines  
\begin{document} \draft

\title{\LARGE \bf Description of Quantum Systems by Random Matrix Ensembles
of High Dimensions\\
"Proceedings of ICSSUR'6, Sixth International Conference on Squeezed States and
Uncertainty Relations, 24th May 1999- 29th May 1999, 
Universit\`{a} degli Studi di Napoli
'Federico II', Naples, Italy", NASA, Greenbelt, Maryland, USA (1999).} 

\author{Maciej M. Duras} 
\address{Institute of Physics, Cracow University of Technology, 
ulica Podchor\c{a}\.zych 1, 30-084 Cracow, Poland\\
AD 1999, July 26th} 
  
\maketitle 

\begin{abstract} 
The new Theorem on location of maximum of probability
density functions of dimensionless second difference
of the three adjacent energy levels
for $N$-dimensional Gaussian orthogonal ensemble GOE($N$),
$N$-dimensional Gaussian unitary ensemble GUE($N$),
$N$-dimensional Gaussian symplectic ensemble GSE($N$),
and Poisson ensemble PE, is formulated:
{\it The probability density functions of the dimensionless second
difference of the three adjacent energy levels
take on maximum at the origin for the following ensembles:
GOE($N$), GUE($N$), GSE($N$), and PE, where $N \geq 3$.}
The notions of 
{\it level homogenization with level clustering}
and
{\it level homogenization with level repulsion}
are introduced.
\end{abstract}

\vspace{8mm}

Many complex $N$-level quantum systems
exhibiting universal behaviour
depending only on symmetry 
of Hamiltonian matrix of the system are divided into:
Gaussian orthogonal ensemble GOE($N$),
or Gaussian unitary ensemble GUE($N$),
or Gaussian symplectic ensemble GSE($N$).
The Gaussian ensembles are used in study of
quantum systems whose 
classical-limit analogs are chaotic.
The Poisson ensemble PE (Poisson random-sequence spectrum)
is composed of 
uncorrelated and randomly distributed energy levels
and it describes quantum systems whose 
classical-limit analogs are integrable.
The standard statistical measure is Wigner's distribution
of the $i$th nearest neighbour spacing:
\begin{equation}
s_{i}=\Delta^{1}E_{i}=E_{i+1}-E_{i}, 
\mbox{\,\,\, i=1,...,N-1}.
\label{ith-spacing}
\end{equation}
For {\it i}th second difference
(the {\it i}th second differential quotient)
of the three adjacent energy levels:
\begin{equation}
\Delta ^{2}E_{i}=\Delta ^{1}E_{i+1} - 
\Delta ^{1}E_{i}=E_{i}+E_{i+2}-2E_{i+1}, 
\mbox{\,\,\, i=1,...,N-2}, 
\label{second difference}  
\end{equation}
we calculated distributions for 
GOE(3), GUE(3), GSE(3), and PE
Refs \cite{Duras 1996 PRE,Duras 1996 thesis,Duras 1999 Phys}.

We formulate the following

\begin{quote}
{\bf Theorem:} {\it
The probability density functions of the dimensionless second
difference of the three adjacent energy levels
take on maximum at the origin for the following ensembles:
GOE($N$), GUE($N$), GSE($N$), and PE, where $N \geq 3$.}
\end{quote}

We present the idea of proof. For Gaussian ensembles
it can be shown that second difference distributions
are symmetrical functions for $N \geq 3$.
Hence, the first derivatives of the distributions
at the origin vanish. For Poisson ensemble
the second difference distribution is Laplace one for $N \geq 3$.
Therefore, the distribution takes on maximum at zero.

The inferences are the following:
\begin{enumerate}

\item The quantum systems show tendency
towards the homogeneity of levels
(equal distance between adjacent levels).
We call it {\it homogeneization of energy levels}.

\item There are two generic homogeneizations:
the first is typical for Gaussian ensembles, the second one for
Poisson ensemble.
For the former ensembles 
we define {\it level homogenization with level repulsion}
as follows.
Energy levels are so distributed that the situation that
both the spacings
and second difference vanish:
\begin{equation}
\Delta^{2}E_{i}=s_{i}=s_{i+1}=0,
\label{second-s1-s2-equal-zero}
\end{equation}
is the most probable one.
For the latter ensemble {\it level homogenization with level clustering}
is described below.
Now it is the most probable
that only the second difference is equal to zero
but the two nearest neighbour spacings are nonzero:
\begin{equation}
\Delta^{2}E_{i}=0, s_{i}=s_{i+1}, s_{i} \neq 0.
\label{s1-s2-equal-zero}
\end{equation}

\item The assumption of non-zero value by the second difference
is less probable than the assumption of zero value.
Equivalently, 
the inequality of the two nearest neighbour spacings
is less probable than their equality.
\item The predictions of the Theorem are corroborated by
numerical and experimental data
Refs \cite{Duras 1996 PRE,Duras 1996 thesis,Duras 1999 Phys}.

\end{enumerate}

The theorem could be extended to other ensembles,
{\it e.g.} circular ones, and it is a direction of future development.

\end{document}